\documentclass[journal]{IEEEtran}

\usepackage{soul}

\usepackage{cite}

\usepackage[rightcaption]{sidecap}

\ifCLASSINFOpdf

  \usepackage{gensymb}

   \usepackage[pdftex]{graphicx}

   \usepackage{multirow}

   \usepackage{tabularew}

   \usepackage[flushleft]{threeparttable}

   \graphicspath{{./pdf/}{./jpeg/}{./eps/}}

  \DeclareGraphicsExtensions{ ,.jpeg,.png}

\else

   \usepackage[dvips]{graphicx}

   \graphicspath{{./eps/}}

   \DeclareGraphicsExtensions{.eps}

\fi

\usepackage{amsmath}

\usepackage{array}

\ifCLASSOPTIONcompsoc

\usepackage[caption=false,font=normalsize,labelfont=sf,textfont=sf]{subfig}

\else

  \usepackage[caption=false,font=footnotesize]{subfig}

\fi

\usepackage{stfloats}

\usepackage{url}

  \usepackage{color}

\usepackage{url}

\begin{document}



\title{Augmented Memory Computing: Dynamically Augmented SRAM Storage for Data Intensive Applications}

\author{Haripriya Sheshadri*, Shwetha Vijayakumar*, Ajey Jacob, Akhilesh~Jaiswal% <-this % stops a space

\thanks{The authors are with the Information Sciences Institute, University of Southern California, Los Angeles, CA-90007, USA}

\thanks{(* These authors contributed equally)}

}


\maketitle

\pagenumbering{gobble}

\begin{abstract}

The emergence of various data-intensive applications like artificial intelligence, machine learning etc., has highlighted the energy and throughput bottlenecks inherent in existing computing systems. Memory storage is a major  component for such data intensive computational platforms. Consequently, memory-centric research investigations to improve the overall energy-efficiency and throughput of a given computing chip is being actively pursued. Exploration of novel memory technologies for high-density on-chip memories and in-memory computing are the key examples of such memory-centric approaches. In this paper, we propose a novel memory-centric scheme based on CMOS SRAM. Our proposal aims at dynamically increasing the on-chip memory storage capacity of SRAM arrays on-demand. The proposed scheme called - \textit{Augmented Memory Computing} allows an SRAM cell to operate in two different modes 1) the Normal mode and  2) the Augmented mode. In the Normal mode of operation, the SRAM cell functions like a standard 6 transistor (6T) SRAM cell, storing one bit of data in static format. While in the Augmented mode, each SRAM cell can store $>$1 bit of data (in a dynamic fashion). Specifically, we propose two novel SRAM cells - an 8 transistor (8T) dual bit storage augmented cell and a 7 transistor (7T) ternary bit storage augmented cell. The proposed 8T dual bit SRAM cell when operated in the Augmented mode, can store a static bit of data while also, simultaneously, storing another bit in a dynamic form. Thus, when operated in Augmented mode, the 8T SRAM cell can store two bits of data - one SRAM-like data and one DRAM-like data, thereby increasing or augmenting the memory storage capacity. On the other hand, the proposed 7T ternary bit storage augmented cell can either store a single SRAM data in Normal mode or can be configured to operate in Augmented mode, wherein it can store ternary data (3 levels (0,0), (0,1), (1,0)) in a dynamic manner. Thus, based on the mode of operation, the proposed augmented memory bit-cells can either store one static bit of data or $>$1 bit of data in a dynamic format. We show the feasibility of our proposed bit-cells through extensive simulations at Globalfoundries 22nm FDX node. It is worth mentioning, the novel scheme of augmented memory bit-cells can be seamlessly combined with existing in-memory computing approaches for added energy and throughput benefits.

\end{abstract}



\IEEEpeerreviewmaketitle

\section{Introduction}

Over decades the advances in hardware computing platforms have been driven by the remarkable scalability of the Metal-Oxide-Semiconductor Field Effect Transistors (MOSFETs) in accordance with Moore's Law \cite{shalf2020future}. Despite consistent improvement in power-performance-area (PPA) metrics, recent trends in data-intensive applications have pushed the state-of-the-art in hardware computing platforms to its limits \cite{gil20201}. Existing hardware solutions suffer from energy and throughput bottlenecks due to frequent data movement between multiple levels of the memory hierarchy and between memory-units and processing-cores \cite{mahapatra1999processor}. To mitigate such bottlenecks various \textit{memory-centric} approaches are being extensively explored by the research community. These include exploration of novel high-density emerging memory technologies \cite{chih202013,wong2010phase} and use of emerging computing approaches like in-memory and near-memory computing \cite{agrawal2018x, sebastian2020memory}. In this paper, we propose a novel memory-centric paradigm - \textit{Augmented Memory Computing} (AMC) for the acceleration of data-intensive applications like artificial intelligence and machine learning. 

It is well-known that the 6 transistor SRAM cell is the most widely used on-chip memory system due to its high robustness and fast read-write speed \cite{kulkarni2020low, song20187nm}. However, a major drawback of SRAM is the associated high area overhead, limiting the amount of on-chip memory storage. Consequently, off-chip memory systems are used as high-density storage at the expense of speed and energy consumption. In fact, the data communication overhead associated with the movement of data from off-chip to on-chip memory forms a major source of energy consumption and compute latency \cite{ahmad2020superslash}. Toward that end, AMC aims at increasing the on-chip storage on demand, thereby dynamically augmenting storage capacity for SRAM arrays that can cater to data intensive applications.

We present a family of novel SRAM bit-cells using two different bit-cell topology - 1) The proposed 8 transistor (8T) augmented SRAM cell includes two additional transistors as compared to a 6T cell, wherein the 8T augmented bit-cell can simultaneously store an SRAM-like and a DRAM-like data based on the applied voltages 2) Similarly, the proposed 7T augmented bit-cell can store a ternary data (-1,0,+1) per bit-cell in a dynamic format as opposed to storing a binary data (0,1) as in a conventional 6T SRAM cell. Note, both the proposed augmented bit-cells can function like normal SRAM cells with comparative read-write margins and speed. As such, our proposed bit-cells can be operated in two distinct modes. In the \textit{Normal} mode these bit-cells function like conventional 6T bit-cell storing a binary data, while in \textit{Augmented} mode, the bit-cells can store more data (either two bits for the 8T augmented cell, and ternary bits for the 7T augmented cell), thereby dynamically increasing the memory storage capacity.
%The resulting increase in on-chip storage capacity reduces the off-chip data access leading to significant improvement in energy and speed of the overall end application. 

 \begin{figure}[t]
    \centering
    \includegraphics[width=0.5\textwidth]{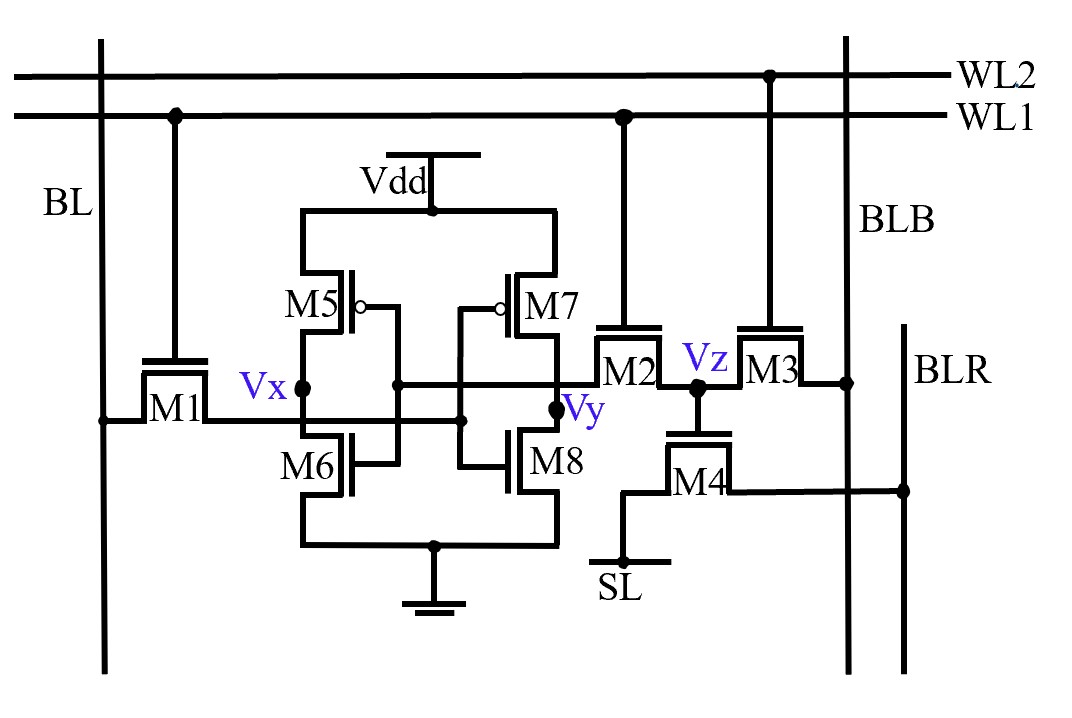}
    \caption{8T dual bit storage augmented bit-cell. The cell can store a SRAM-like and a DRAM-like date, simultaneously, in the augmented mode.}
    \label{fig1}
\end{figure}

It is worth mentioning, in comparison  to in-memory computing, AMC does not rely on complicated approximate analog computing and hence is more robust. 
%Furthermore, AMC can be integrated seamlessly with standard von-Neumann systems without the need to revamp existing computing architectures for enabling radically different paradigms like in-memory computing.
This does not imply that AMC cannot be used in conjunction with in-memory computing. As discussed in section IV, AMC can be combined with existing in-memory computing approaches for improved energy-efficiency and throughput \cite{agrawal2018x, 8tdpe, si2019dual}. In summary, although AMC is conceptually independent of in-memory computing paradigms, yet it can be easily combined with existing in-memory processing schemes. Thus, AMC presents a novel approach for memory-centric computing, along with other existing memory-centric approaches (like in-memory/near-memory computing). The key contributions of the paper are mentioned below:

\begin{figure*}[t]
    \centering
    \includegraphics[width= .9\textwidth]{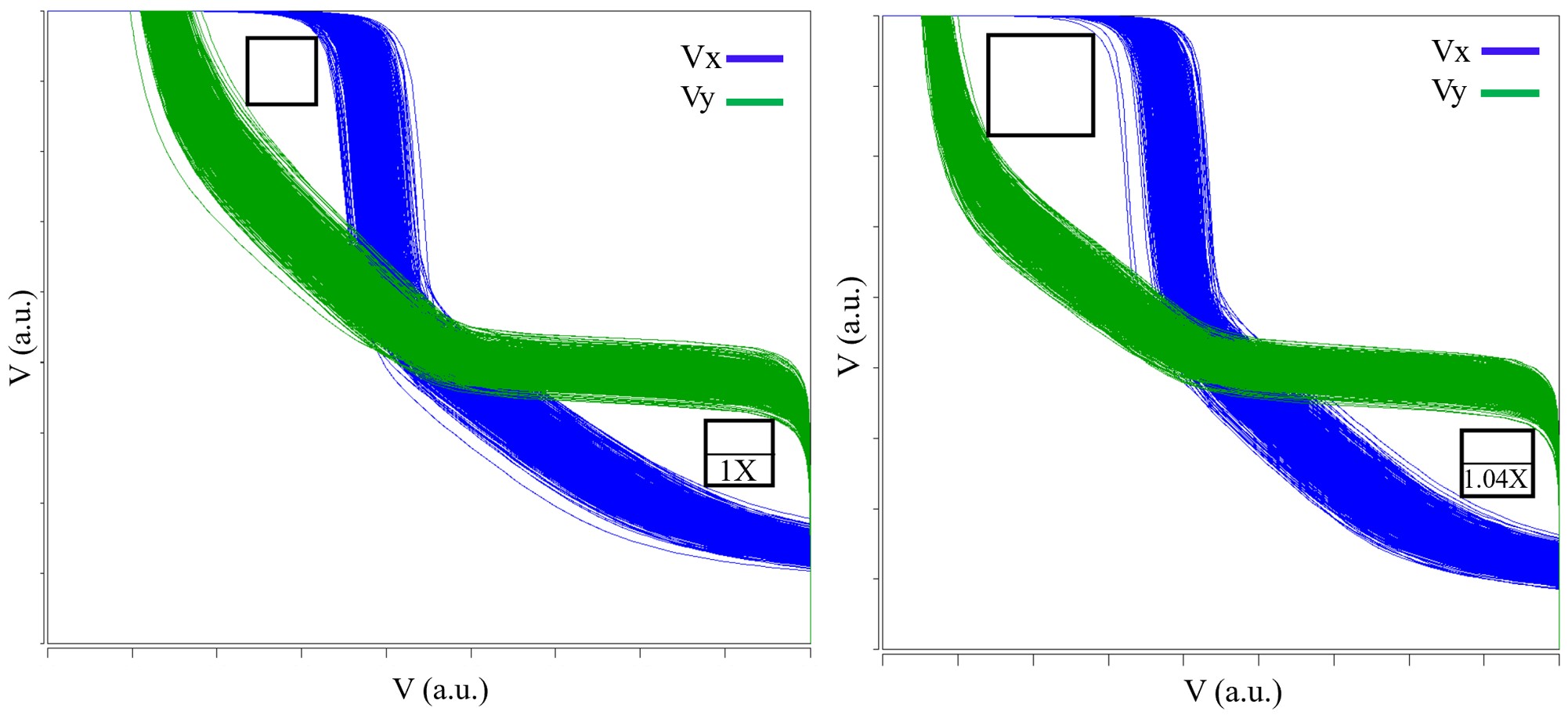}
    \caption{(Left) 6 transistor SRAM cell read SNM using SRAM bit-cell from Globalfoundries 22nm library. (Right) The read SNM for the proposed 8 transistor augmented bit-cell. As seen the read SNM for the proposed bit-cells is similar to the library SRAM bit-cell, thereby demonstrating the asymmetry in the SRAM circuit does not drastically alter the cell stability.}
    \label{fig3}
\end{figure*}

\begin{enumerate}

\item We propose a novel memory-centric compute paradigm called \textit{Augmented Memory Computing}, wherein the storage capacity of on-chip SRAM based memory can be increased dynamically for data intensive computations.
%\item We exploit the data intensive nature of emerging applications like AI and ML to our advantage, wherein with increased on-chip storage despite reduced robustness the overall system level energy and throughput can be improved. This improvement results due to less movement of data to external off-chip storage. 
\item We specifically propose two distinct bit-cells. The proposed 8T augmented bit-cell can store two bits of data, simultaneously, (an SRAM-like and a DRAM-like data), while the proposed 7T augmented bit-cell can store ternary data (+1,0.-1) in a dynamic fashion. \item The proposed augmented bit-cells can be used either in Normal mode or Augmented mode. Furthermore, our proposed 8T augmented bit-cell shows better voltage scalability as compared to 6T SRAM cell in Normal mode, similarly the 7T augmented bit-cell allows fine-grained power gating compared to 6T bit-cell in Normal mode.
\item We present detailed circuit simulations reporting energy, speed metrics along with variation analysis at Globalfoundries 22FDX (22nm FD-SOI). \cite{22fdxx} node %along with system-level energy and throughput estimates for running a deep neural network using the proposed augmented memory computing paradigm.

\end{enumerate}

The rest of the paper is organized as follows. In section II, we present the 8 transistor augmented bit-cell. We first discuss the Normal mode of operation followed by the dual bit storage Augmented mode of operation. Section III presents a 7 transistor augmented bit-cell, that can store a ternary bit within each bit-cell in the Augmented mode of operation. Section IV presents simulation results and discussions regarding the use of augmented bit-cells in conjunction with in-memory computing. Finally, section V concludes the paper.

\section{8T Dual Storage Augmented Bit-Cell}

As mentioned earlier, augmented bit-cells can increase their storage capacity dynamically while also functioning like conventional SRAM bit-cells in the Normal mode. For a dual bit augmented storage, we add two additional transistors to the 6T SRAM cell, as shown in Fig. \ref{fig1}.  This bit-cell can operate in two distinct modes - the Normal mode and the Augmented mode. The SRAM cell can be configured in the Normal or Augmented mode at a sub-array level granularity.

 \subsection{Normal Mode}
 
 For the Normal mode of operation, both the wordlines WL1 and WL2 are activated, simultaneously, during the read and write operations. The resulting SRAM read and write operations are similar to the 6T cell, except that the SRAM is asymmetric due to the presence of an additional access transistor (M3) on the BLB side of the 8T SRAM cell. As shown in Fig. \ref{fig3}, using 1000 Monte-Carlo simulations at GF 22FDX node, we observe almost no change in static noise margins, compared to the 6T bit-cell. Thus, the asymmetric nature of the 8T bit-cell leads to minimal alteration of static noise margins and hence the cell stability. Additionally, the SL line and the BLR line are kept at 0V during the Normal mode of operation. This ensures no current flows through transistor M4, irrespective of the voltage at its gate at the node Vz. In summary, the 8T bit-cell shown in Fig. \ref{fig1} can be operated similar in functionality to the conventional 6T SRAM cell when both WL1 and WL2 are simultaneously activated. This also implies that a conventional differential sense-amplifier can be used to sense the data stored in the SRAM cell.

Advantageously, the presented 8T bit-cell can be used to improve cell-stability and achieve lower operating voltages compared to the 6T bit-cell using the well-known pulsed wordline activation scheme \cite{khellah2006wordline}. This can be achieved by using transistor M4 as a de-coupled read port. For improved stability, WL1 is activated first using a short duration pulse, keeping WL2 OFF. This would ensure node Vz is charged or discharged based on the data stored at node Vy. Essentially, by pulsed activation of the wordline WL1, we are copying the SRAM data into the node Vz. Since the pulse duration of the signal on WL1 would be much smaller than conventional 6T SRAM the possibility of read disturb is minimal \cite{khellah2006wordline}. After the pulsed activation of WL1,  the SL is pulled to $V_{DD}$. Subsequently, a large signal inverter-based sensing can detect the voltage change on BLB. Thus, the proposed 8T bit-cell can be used in conjunction with a pulsed WL scheme to improve cell-stability.

 \begin{figure}[t]
    \centering
    \includegraphics[width=0.5\textwidth]{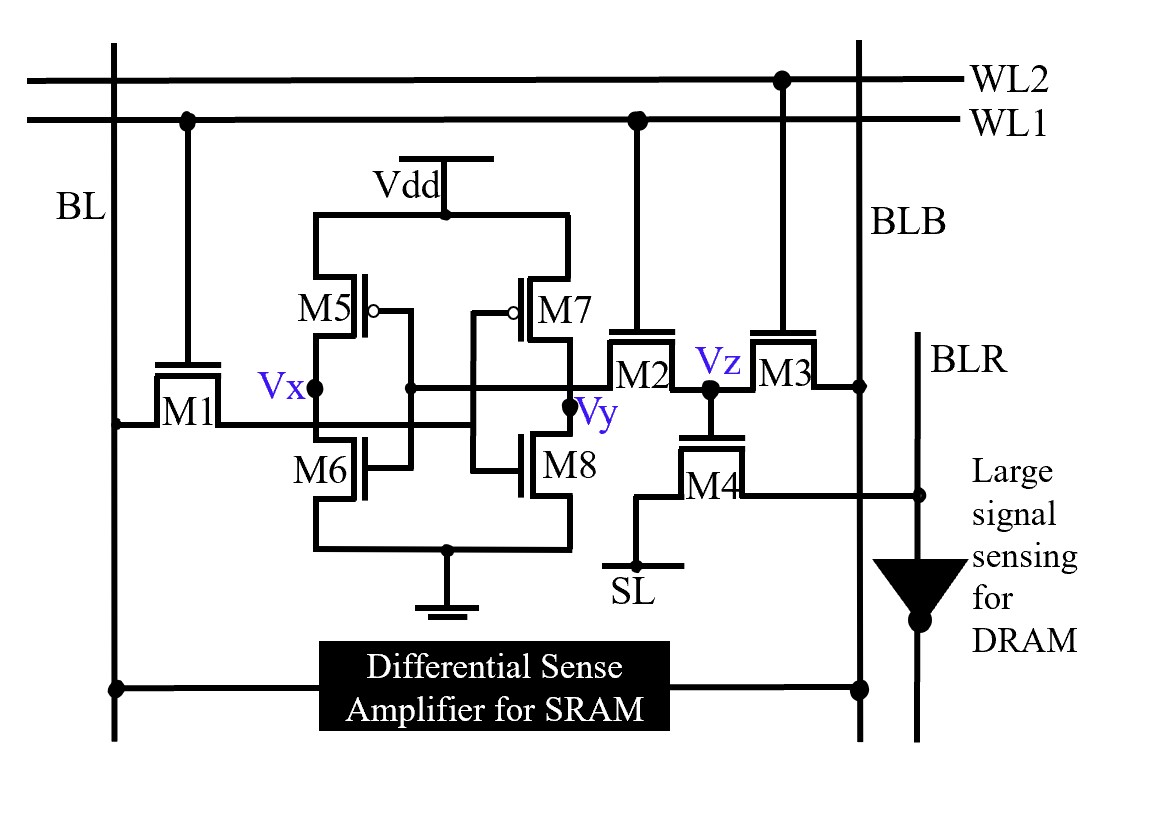}
    \caption{Figure showing the sensing scheme for the SRAM and DRAM data in the augmented 8T bit-cell. The SRAM data can be sensed through the differential bit-lines BL and BLB, while the DRAM data can be sensed using single ended large signal inverter based sensing. Note, a FILO scheme ensures the DRAM data is not disturbed inadvertently.}
    \label{figx}
\end{figure}

\subsection{Augmented Mode}

In the augmented mode, the 8T bit-cell stores two bits of data, simultaneously. The SRAM-like static data is stored in the cross-coupled inverter pair as complementary voltages on nodes Vx and Vy, similar to the conventional 6T SRAM storage; while the two transistors M3 and M4 store a DRAM-like data on the dynamic node Vz. In fact, transistors M3 and M4 form a 2 transistor embedded DRAM cell, which can be written by activating WL2, and can be read through transistor M4 using lines SL and BLR. For the DRAM write operation, line WL2 is pulled high and data is written into the DRAM node Vz through the line BLB. Note, due to the presence of an NMOS only access transistor M3, we use voltage boosting on WL2 for writing a high value at the dynamic node Vz. During the write operation, the SL lines are all kept at 0V and the BLR lines are also discharged to 0V. For the DRAM read operation, the corresponding SL line is pulled high and a voltage accumulation on the initially discharged line BLR is sensed to read the DRAM data. Note, all unselected SL lines are kept at 0V. The DRAM data can be read by using a large signal inverter based sensing as shown in Fig. \ref{figx}. The compact inverter based sensing ensures minimal sensing circuit overhead.  In summary, the transistors M3, M4 along with lines SL and BLR constitute an embedded DRAM cell within the 8T SRAM cell, such that it can store an independent data in a dynamic fashion, while simultaneously a static data is stored in the SRAM cell. 

Interestingly, while transistors M3, M4 and node Vz store a DRAM-like data, the SRAM data is stored on nodes Vx and Vy which can be read and written by simultaneously activating worldines WL1 and WL2. The SRAM data can be read using a latch based differential current or voltage sense amplifier, as shown in Fig. \ref{figx}. It is important to note that the DRAM data stored on node Vz will be destroyed during the read or write operation of the SRAM data since the DRAM node Vz is in the SRAM read/write path. However, this issue can be circumvented by relying on the data access pattern for specific end applications. For example, in a deep learning network nodes Vx and Vy can store weights while the corresponding node Vz streams input activations. %Thus, for computing purposes activation will first be written into the SRAM part of the 8T cell followed by the kernels in the DRAM part of the 8T cell. For computations, the weight kernels will be first read from the DRAM cell followed by the activation from the SRAM part.
In other words, by ensuring a first-in last-out (FILO) scheme for the combined SRAM-like and DRAM-like bit we can store two bits simultaneously in the 8T bit-cell without inadvertently destroying the DRAM data. Given the regular memory access pattern for data intensive applications like artificial intelligence and machine learning, it is easy to enforce such a FILO scheme. 

\begin{figure}[t]
    \centering
    \includegraphics[width=0.47\textwidth]{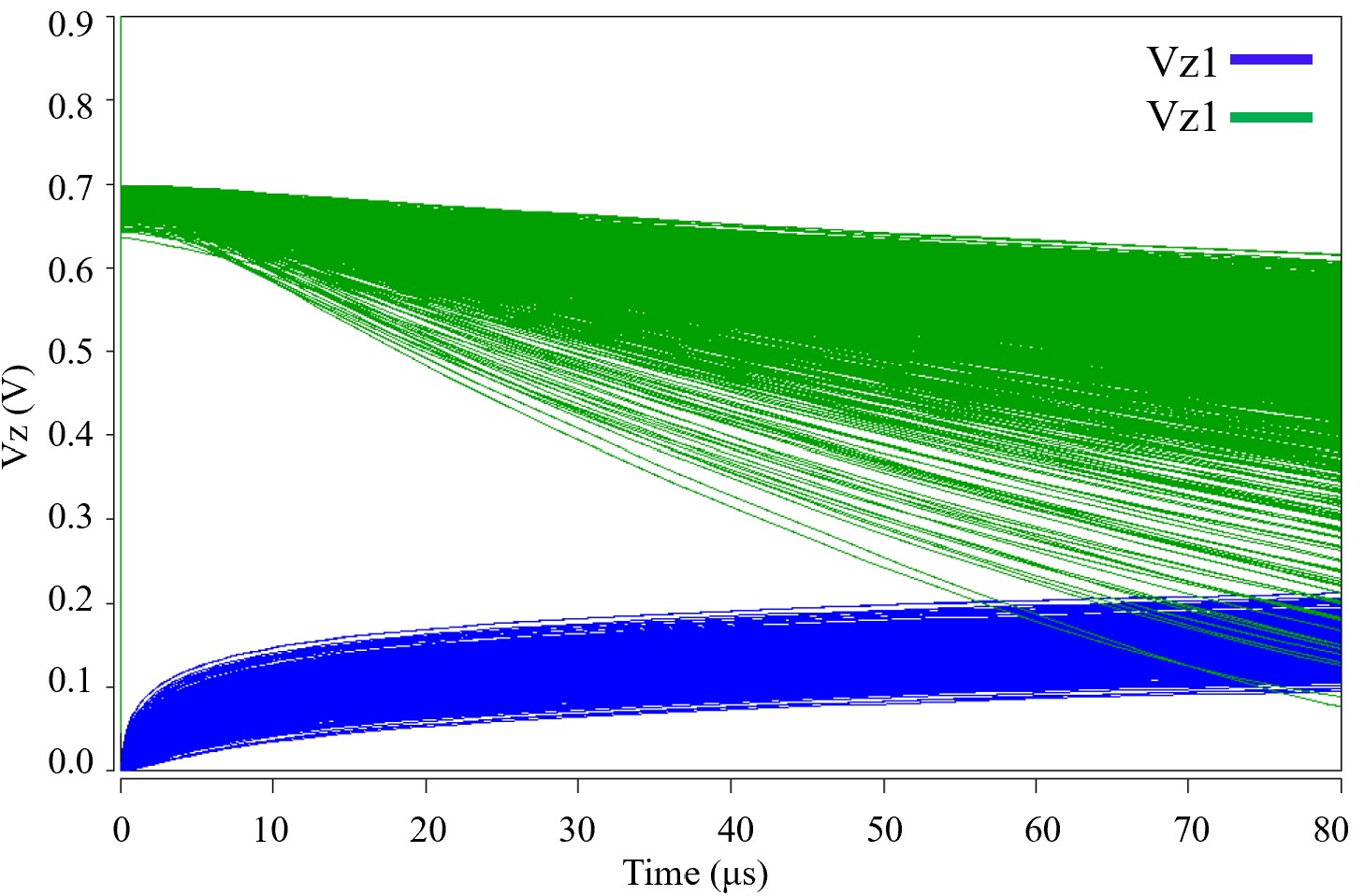}
    \caption{Figure depicting leakage of data stored on the dynamic node as a function of time for the DRAM-like bit storage in the 8T augmented cell at 85C.}
    \label{fig4}
\end{figure}

%Furthermore, the DRAM data requires periodic refresh to ensure the stored data on the dynamic node XX does not leak away. For the case of accelerating a deep network wherein activations are stored in the SRAM bit-cell and weights are streamed through the DRAM, the weights are to be read only once for a given activation. As such, if the DRAM retention time is sufficient to allow data storage for a time longer than the maximum compute time for a particular slice of input activation the DRAM data would not be needed once it is read and a convolution operation has already been performed with corresponding input activations. 
Thanks to the excellent leakage control of the 22nm GF FD-SOI technology, we have achieved 25$\mu$s retention time at 85C with a small (-100mV negative voltage) on wordlines WL2 during hold mode of the memory array. Timing waveforms with Monte-Carlo simulations are shown in Fig. \ref{fig4},  exhibiting the retention waveform at a temperature of 85 degrees Celsius. The retention time is defined here as the time until the peripheral circuit can reliably sense the data stored on the dynamic node Vz.% Table I provides a summary of retention time under various conditions. Note, the retention time can be improved using various design knobs that include use of high-VT transistors, body biasing \cite{narinx201924} etc. %In summary, the proposed 8T cell during Augmented Mode can store a SRAM-like and a DRAM-like data. Data integrity is ensured by enforcing a LIFO scheme for the bit-cell, while the need for a refresh operation is obviated through use of low leakage FD-SOI technology coupled the data access pattern for a typical deep learning network.

\section{7T Ternary Storage Augmented Bit-Cell}

It is well-known that a 6 transistor SRAM cell can store one digital bit in a static format. SRAM being a differential memory, both the bit and the complement of the bit are stored in the same cell. Our proposed 7T augmented SRAM cell can either be configured to store one static SRAM bit (Two Levels: Normal mode of operation) or one dynamic ternary bit (Three Levels: Augmented mode of operation). Note, ternary bits have three levels usually represented as (-1, 0, +1). 

Ternary memory storage is becoming increasingly popular due to the recent algorithmic advances in Ternary Neural Networks (TNNs). TNNs are being extensively explored \cite{alemdar2017ternary, mellempudi2017ternary}, since they provide both lower memory requirement as well as improved accuracy for deep learning networks. Traditionally, since 6T SRAM cell can only store a binary data, two 6T cells are required to store one ternary data \cite{jain2020tim}.  As such, our proposed 7T Ternary augmented cell can be configured to increase the on-chip SRAM storage density for ternary weights for TNN accelerators. Note, the proposed 7T Ternary augmented cell stores ternary data in a dynamic format as opposed to the conventional 6T SRAM, which stores static binary data. %Nevertheless, significant energy and through improvements can be obtained using the proposed AMC, thanks to the characteristic data access pattern for a deep learning network.

 \subsection{Normal Mode}
\begin{figure*}[t]
    \centering
    \includegraphics[width=0.8\textwidth]{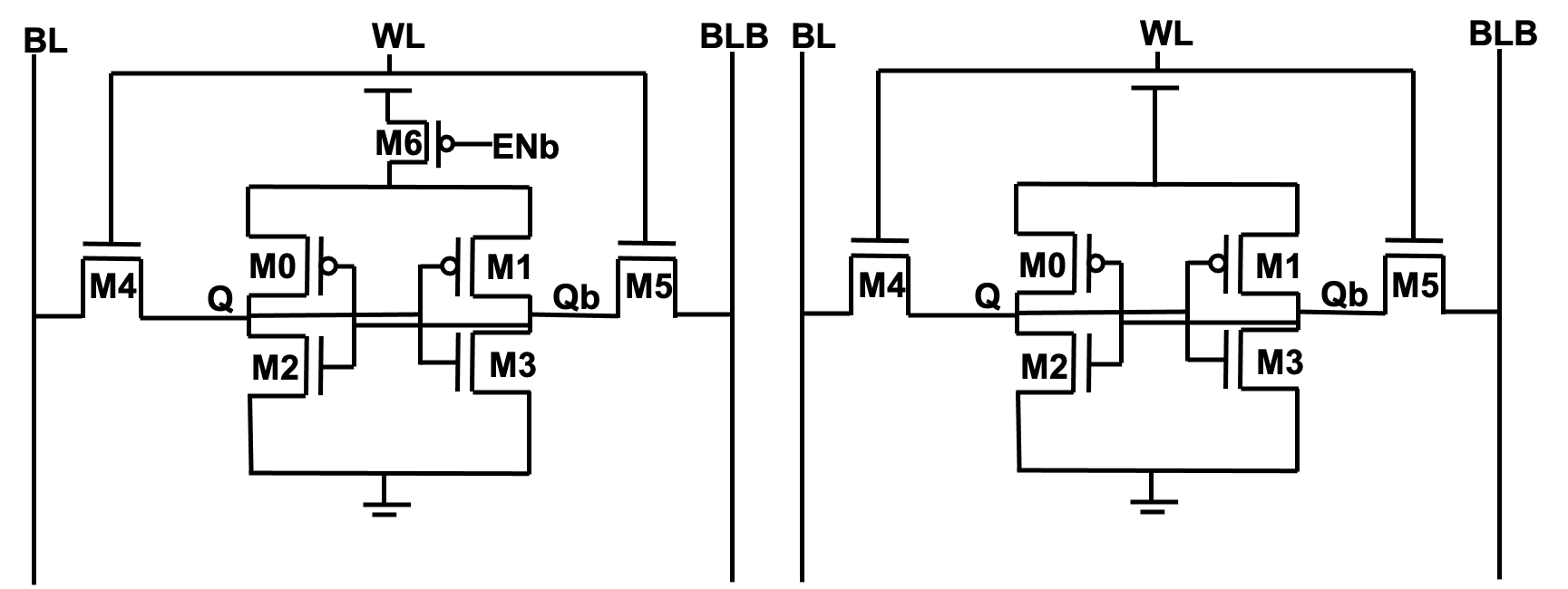}
    \caption{7T augmented ternary bit-cell in Normal mode. During the normal mode of operation M6 is ON, thereby the 7T cell acts like a conventional 6T SRAM cell.}
    \label{fig:figure1}
\end{figure*}

 The 7T Ternary augmented cell is shown in Fig. \ref{fig:figure1}. It consists of 6T SRAM cell with one additional PMOS transistor per bit-cell connecting the cross-coupled inverters to VDD. For the Normal mode of operation, the PMOS M6 is kept ON, and the augmented cell functions similar to a normal SRAM cell. It is worth mentioning, that similar 7T cells with a header PMOS have been used in previous literature to enable fine-grained power gating \cite{powell2000gated}. In the traditional use-case, when the PMOS transistor M6 is switched OFF, the cross-coupled inverters are disconnected from the power supply and the SRAM cell enters power-gated mode, thereby drastically reducing the cell leakage. In summary, the normal mode of operation of the 7T augmented cell is similar to a gated VDD SRAM cell \cite{powell2000gated}. When PMOS M6 is ON, the cell stores one static data, when PMOS M6 is OFF, the cell is disconnected from VDD and is in power gated mode.

 \subsection{Augmented Mode}
 
\begin{figure}[b]
    \centering
    \includegraphics[width=0.4\textwidth]{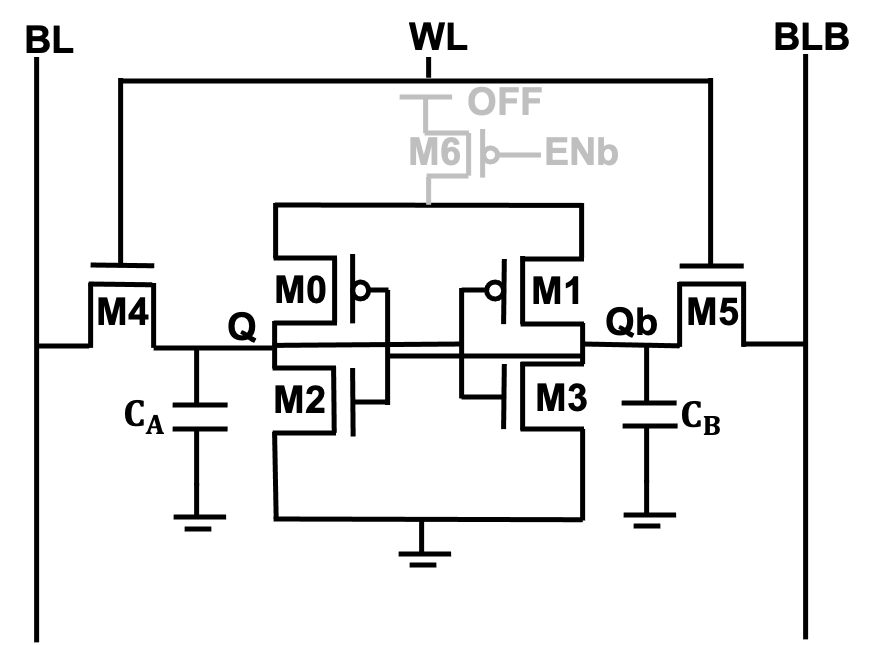}
    \caption{7T augmented cell showing parasitic capacitances at nodes Q and QB that act like dynamic nodes to store the ternary data (0,1), (1,0) or (0,0) in a DRAM-like fashion during augmented mode of operation. }
    \label{figw}
\end{figure}

During the Augmented mode, transistor M6 is switched OFF. As seen in Fig. \ref{figw}, the cross-coupled inverters are disconnected from VDD, and the SRAM can no longer store data and its complement in static format. With PMOS M6 OFF, since the cross-coupled inverters are no longer connected to VDD, the positive feedback behavior of the cross-coupled inverters are weakened. We exploit this weakened feedback between the nodes Q and QB to store three different data patterns on nodes (Q,QB) in a dynamic fashion. Specifically, with the PMOS M6 OFF, (Q,QB) can dynamically store one among three different data patterns $-$ (1,0), (0,1) and (0,0). This three level storage allows us to store a ternary bit in the 7T SRAM cell, wherein (1,0), (0,1) and (0,0) can be conceptually mapped to ternary levels of (-1,+1, 0), respectively. 

\begin{figure}[t]
    \centering
    \includegraphics[width=0.5\textwidth]{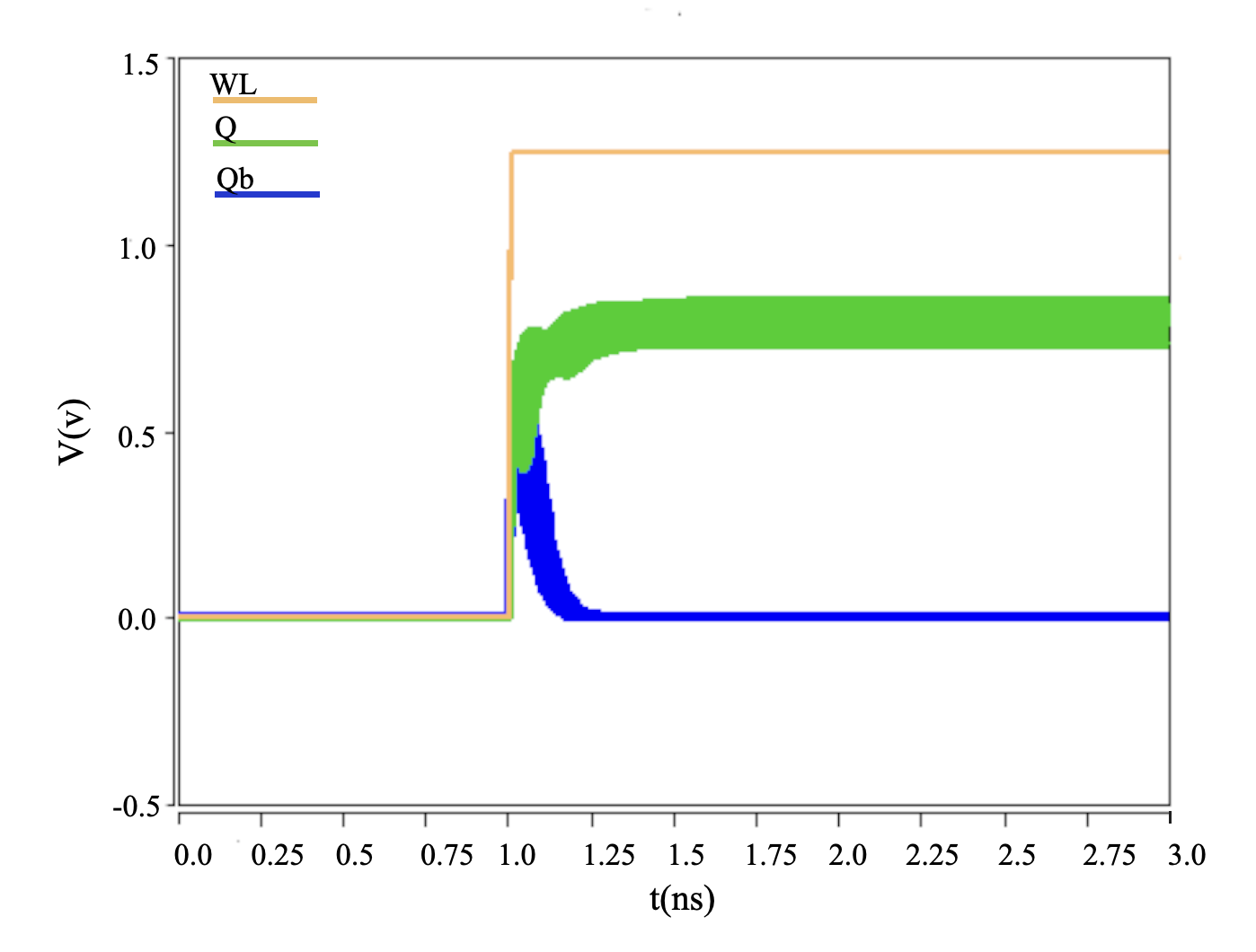}
    \vspace{.001in}
    \caption{Waveforms showing the writing of data (0,1) / data (1,0) into the augmented 7T SRAM cell when PMOS M6 is OFF.}
    
    \label{figwx}
\end{figure}

The storage of different voltage levels on nodes Q and QB can be understood as follows. Let us consider the case of storing the data (0,1) in the SRAM cell during the Augmented mode (i.e. when PMOS M6 is OFF). Keeping PMOS M6 OFF, the WL is activated, BL and BLB are pulled high or low based on the data to be stored in the SRAM cell. Note, since the SRAM cell is not connected to VDD, the cross-coupled inverter does not have a positive feedback mechanism to write a robust `1' into the cell. This is because the access transistors consisting of NMOS devices suffer a threshold voltage (VT) drop when passing a high value. Consequently, we use voltage boosting, wherein the WL is pulled to 1.25V to ensure a strong `1' is written into the SRAM cell. The write waveforms for storing data (0,1) is shown in Fig. \ref{figwx}. Note, with PMOS M6 OFF, when Q and QB stores (0,1), the DRAM-like capacitors $C_{A}$ and $C_{B}$ (formed due to parasitic gate and diffusion capacitances) are charged and discharged, respectively (see Fig. \ref{figw}). As a result, transistors M1 and M2 are ON and M0 and M3 are OFF. By symmetry, the waveforms of Fig. \ref{figwx} also represent write operation for data (1,0) with appropriate voltages being applied on BL and BLB. 

\begin{figure}[t]
    \centering
    \includegraphics[width=0.5\textwidth]{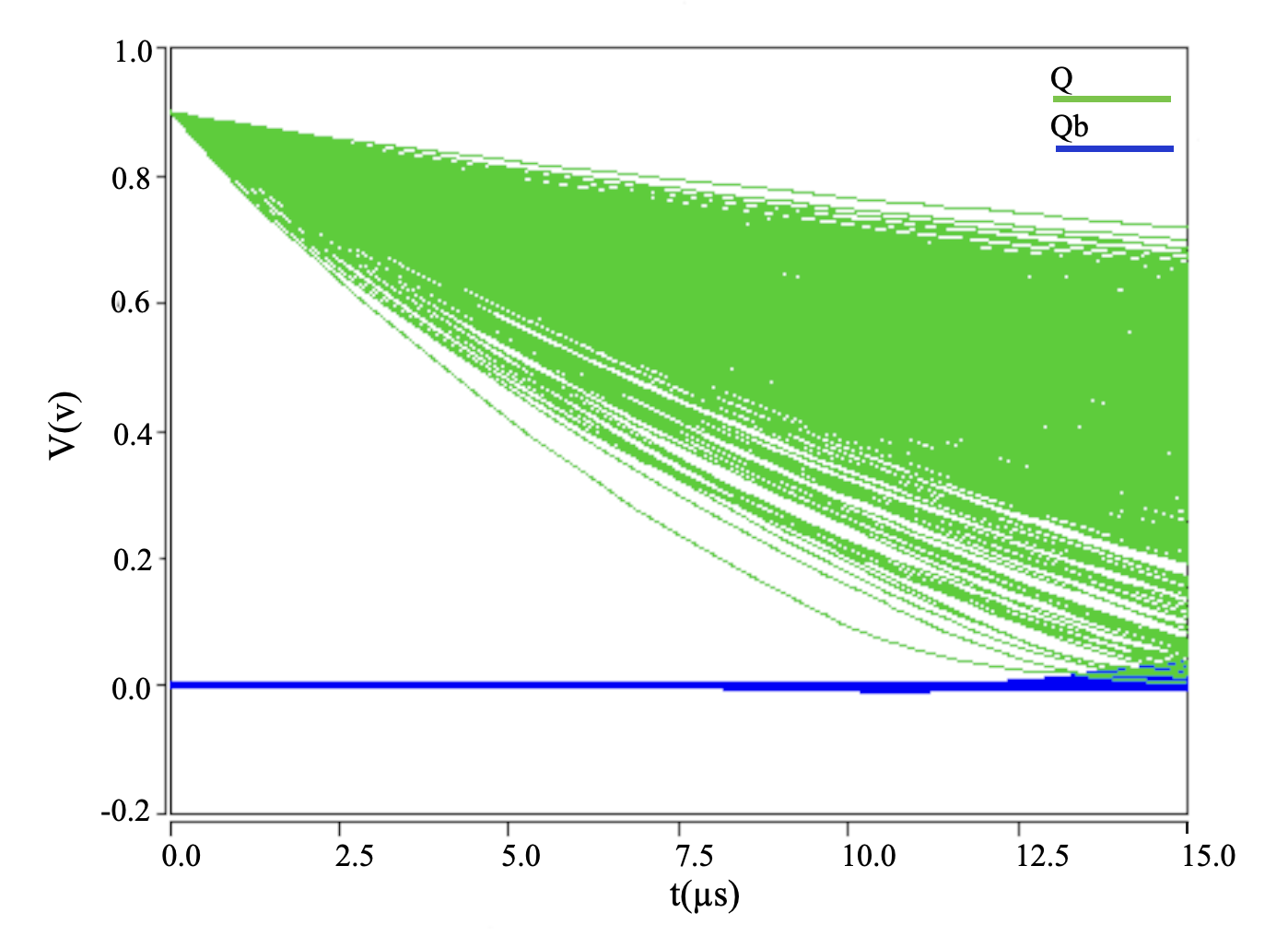}
  %  \vspace{.01in}
    \caption{Data retention time for data (0,1) / data (1,0) at 85C. The retention time improves with decreasing temperature.}
    \label{cccc}
\end{figure}

As expected, due to absence of a VDD connection, over a certain duration of time, the dynamic nodes Q and QB leak, destroying the data stored in the augmented cell. Thereby, with respect to (0,1) data, the retention time can be defined as the time up to which the data Q and QB can be robustly sensed by the peripheral sensing circuit. A similar argument for retention time can also be made for data (1,0). Figure showing the leakage of voltages on nodes Q and QB that dictates the retention time  for data (0,1)/(1,0) is depicted in Fig. \ref{cccc}.

\begin{figure}[t]
    \centering
    \includegraphics[width=0.55\textwidth]{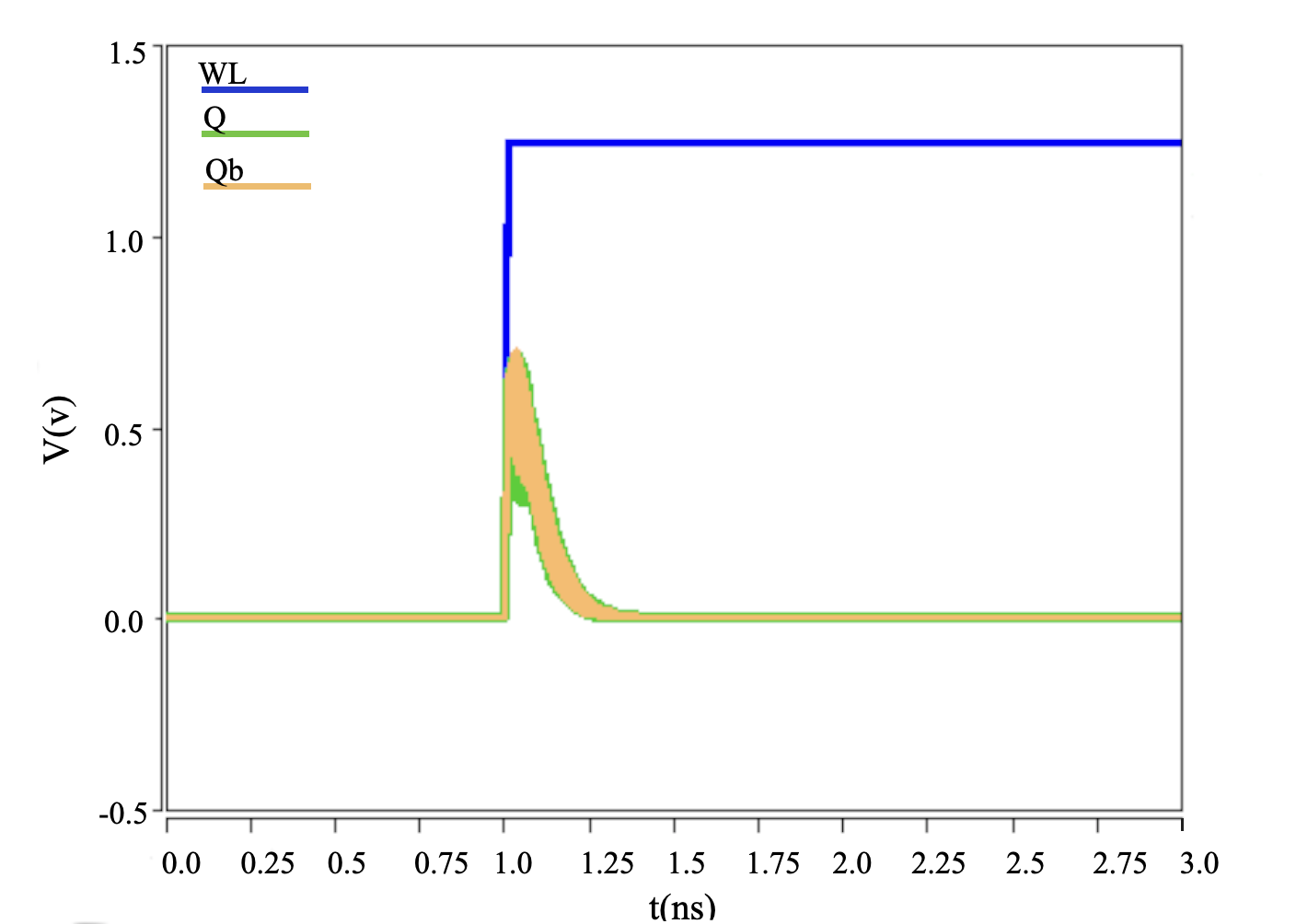}
     \caption{The writing of data (0,0) in the 7T augmented cell. In absence of positive feedback, WL can be activated and BL and BLB can be pulled low to write data (0,0).}
    \label{fig00}
\end{figure}

Now let us consider storing data (0,0) in the augmented 7T cell. For writing the data (0,0), both Q and QB are discharged by activating the WL and pulling BL and BLB to 0V. This in turn switches OFF NMOS transistors M2 and M3. Although the PMOSes M0 and M1 are ON, nodes Q and QB are disconnected from VDD due to the OFF transistor M6. As a result, nodes Q and QB are not  connected to VDD. The capacitors $C_A$ and $C_B$ act as dynamic floating nodes that are neither connected to GND nor to VDD. Thus, the dynamic nodes Q and QB store the data (0,0) when BL and BLB are pulled low and WL is high. The write waveforms for data (0,0) is shown in Fig. \ref{fig00}. Note, in a conventional 6T SRAM cell, storage of (0,0) is not possible due to the existence of strong positive feedback between the two cross-coupled inverters. The strong positive feedback forces Q and QB to always be the complement of each other. By disconnecting the VDD using PMOS M6, we significantly weaken the feedback connection and store (0,0) on nodes (Q,QB) on the dynamic capacitors $C_A$ and $C_B$. In summary, when PMOS M6 is switched OFF, the 7T cell can either store (1,0), (0,1), or (0,0) in a dynamic manner. Note, as expected the data (0,0) stored on dynamic nodes Q and QB leak with time. However, the resulting retention time is higher than the case of (0,1) and (1,0). Consequently, the (0,0) data storage does not decide the overall retention of the 7T augmented cell, it is rather limited by the retention time of data (0,1)/(1,0) as shown in Fig. \ref{cccc}.

%\begin{figure}[t]
%    \centering
%    \includegraphics[width=0.5%\textwidth]{read_01_10.png}
%    \caption{(a) For Normal mode of operation a differential sense amplifier can be used for sensing (b) For the case of Augmented mode two inverters along with a digital logic circuit is employed to sense ternary data (0,1), (1,0) and (0,0).}
 %   \label{fig:figure6}
%\end{figure}

\begin{figure}[t]
    \centering
    \includegraphics[width=0.5\textwidth, height= 2.2in]{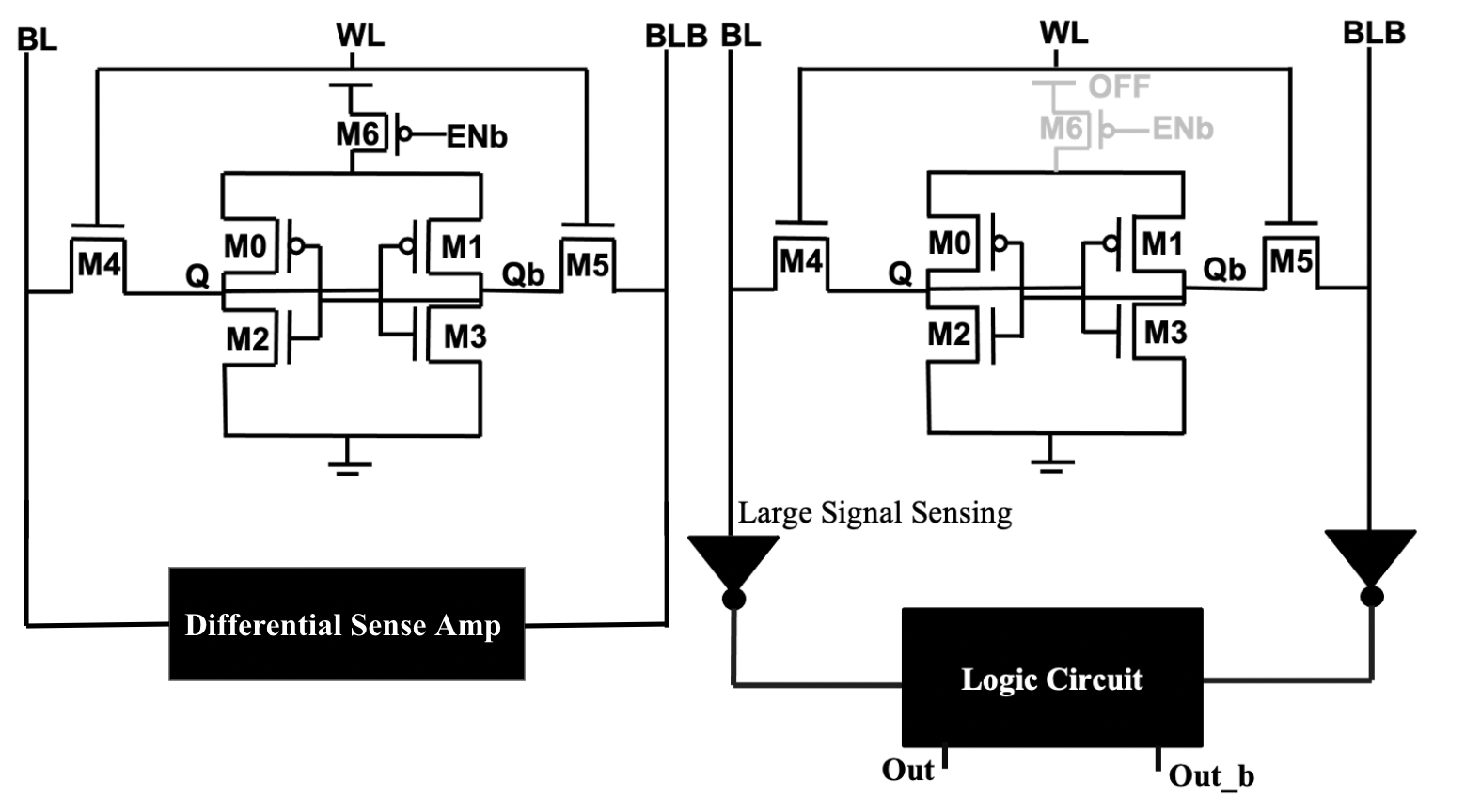}
    \caption{(a) For Normal mode of operation a differential sense amplifier can be used for sensing (b) For the case of Augmented mode two inverters along with a digital logic circuit is employed to sense ternary data (0,1), (1,0) and (0,0).}
    \label{figamc}
\end{figure}

\begin{figure}[t]
    \centering
    \includegraphics[width=0.5\textwidth]{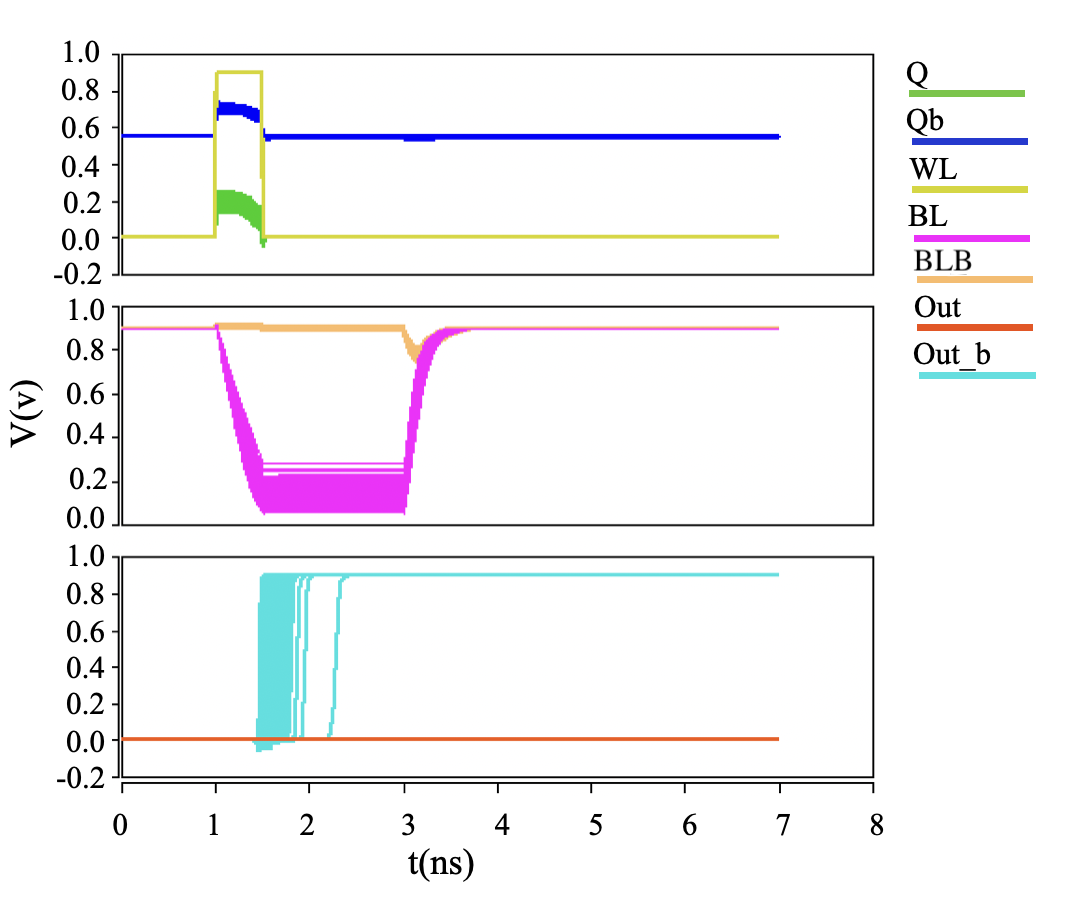}
    \caption{Read waveform for data (0,1) at the end of retention time at 85C.}
    \label{dddd}
\end{figure}

Let us now consider the readout of data (0,1), (1,0) and (0,0). For a (0,1) or (1,0) readout, one of the NMOS transistors M2 or M3 is ON depending on the data stored in the bit-cell. When WL is made ON by pulling it to VDD, the pre-charged BL or BLB discharges depending on which among NMOS transistors M2 or M3 is ON. Thus, by sensing the discharge on BL and BLB, the peripheral circuit can conclude if the data stored in the augmented 7T cell is (0,1) and (1,0), respectively. In its simplest form, the readout circuit consists of large-signal inverter-based sensing, as shown in Fig. \ref{figamc} (b). On the other hand, for the data (0,0), both the NMOS transistors are OFF and both BL and BLB do not see a significant discharge except for leakage currents. As such, no significant voltage discharge either on BL or BLB indicates storage of data (0,0). In summary, a discharging BL indicates data (0,1), a discharging BLB indicates storage of data (1,0), and no significant discharge either on BL or BLB indicates data (0,0). The logic circuit shown in Fig \ref{figamc}(b) takes the voltage output from the sensing inverters as input digital signal and converts it into (0,1), (1,0), or (0,0) representing the data stored in the 7T SRAM cell. Read waveforms for reading the data (0,1) is shown in Fig. \ref{dddd}. By symmetry, the waveforms also represent the readout of data (1,0). Note, in the case of (0, 0), both BL and BLB would not show any significant discharge during the read operation, as such, the waveforms for (0, 0) read are not shown explicitly in the figure. It is worth mentioning, during Normal SRAM operation i.e. when PMOS transistor M6 is ON the 7T SRAM cell functions like a standard 6T cell that can be sensed using standard differential sensing amplifier. In addition, one could also use two differential sense amplifiers, one on each BL and BLB for small signal single ended sensing instead of large signal sensing that uses inverters. Although, such differential sensing during augmented mode has speed benefits, it suffers from area overhead drawback.

%\begin{figure}[h]
 %   \centering
  %  \includegraphics[width=0.5\textwidth]{retention_00.png}
   % \caption{Data Rentention time for data (0,0)}
    %\label{figret}
%\end{figure}

\section{Results and Discussions}

As discussed earlier, the proposed augmented bit-cells can be operated in two modes - the Normal mode and the Augmented mode. In essence, augmented cells rely on dynamic storage within the SRAM cells to increase the memory storage capacity. Due to such dynamic nature of storage, retention time is the key metric for augmented bit-cells. Furthermore, the retention time shows a strong dependence on temperature and makes our proposed cells interesting for cryo-computing applications \cite{cryo}. Table \ref{tab1}-\ref{tab2} mentions the retention time for various temperatures. The retention time is a strong function of temperature and are in similar range as reported in previous works on embedded-DRAM cells \cite{giterman20174}. The retention time can be improved using  circuit based design knobs like body biasing \cite{narinx201924} etc. In addition, based on the end application requirement, a hardware-algorithm co-design approach can be used to allow relaxed retention times by leveraging the resiliency of the end application, for example, the resilient nature of a deep learning network can be used to extend the retention time of the augmented bit-cells using error-aware training of the neural network.

In Table \ref{tab3}-\ref{tab4} we have enumerated read, write energy associated with the augmented cells in Normal and Augmented modes along with leakage power consumption. For the sake of comparison, we have also mentioned the energy and delay numbers for conventional 6T SRAM cell at 22nm Globalfoundries FDX technology. All the energy and power numbers are reported for operation at a temperature of 85C. For the 8T augmented cell, the read, write energy and leakage power increases as compared to a 6T cell. This increase can be attributed to more number of transistors in the augmented cell, that add both  parasitic capacitance and leakage energy consumption. Also, it was observed due to the use of single ended sensing for the DRAM part of the 8T augmented cell the read energy increases by 2.7x compared to the 6T cell. For the 7T augmented cell, the energy metrics are comparable to the 6T SRAM cell for Normal mode of operation. The reduced write energy can be attributed to the OFF PMOS header transistor in the augmented mode making write operation easier and reducing cell leakage for unselected rows. Table \ref{tab5}-\ref{tab6} report the read and write time for the augmented mode operation. Note, due to the presence of BL and BLB the read delay for 7T bit-cell is lesser as compared to the 8T bit-cell. The delay number for 8T bit-cell is for the DRAM-like bit, the SRAM-like bit storage has similar read, write dealy as a normal 6T SRAM cell.

%A key metric for the proposed bit-cells in augmented computing mode is data retention time. 

%%In order to estimate the energy and throughput benefit of the proposed AMC bit-cells, we take example of two networks (a 4 bit quantized 50 Layer Network [Ref] and a Ternary Weight Network [Ref]) that have been shown to provide state-of-the-art accuracy on standard datasets. The details of these networks are summarized in Table XX. Further, we consider a generic architecture consisting of external DRAM, on-chip SRAM and multiple processing engines (PE). Each PE is assumed to be capable of computing sum pf products as required for various neural networks. Given the limited on-chip SRAM size, only apart of the weights and activation per layer can be stored in the SRAM memory. As a result, the data has to be frequently shuttled back and forth between the external DRAM and the on-chip SRAM.
%The energy and latency for a given layer can be estimated by using equations XX and YY. Details of various parameters used in the equations are summarized in Table XX. It can be observed that the external DRAM access is the major contributor for both energy and latency for the neural networks under consideration. 

%To estimate the energy and latency benefits of the AMC bit-cells, we re-evaluate the energy and delay number, considering the on-chip SRAM is now replaced by AMC bit-cells. For iso-SRAM area, replacing 6T SRAM with 8T dual bit AMC cell allows ....

Let us now highlight some key discussions with respect to the augmented memory bit-cells. As detailed in the manuscript, augmented memory bit-cells bring in a novel approach to dynamically increase the memory storage capacity. As such, the augmented bit-cells help to alleviate the issues associated with limited on-chip storage. On the other hand, in-memory computing is another well-known approach being extensively investigated by the research community \cite{sebastian2020memory, 8tdpe, agrawal2017x}. Below are the key points we would like to highlight about augmented memory with respect to in-memory computing.

AMC aims at dynamically increasing on-chip storage capacity through modified SRAM bit-cells. It is important to note that, use of augmented mode does not incur any approximation in data storage or computed data. The sole difference between normal SRAM and augmented storage is the dynamic nature of data and does not effect the accuracy of computations. This is in contrast to in-memory computing paradigms, wherein multiple rows are activated and computations are achieved through approximation of the accumulated signal on the bit-lines. Thus, AMC paradigm is more amenable to traditional memory verification and design flow than in-memory paradigms.

\begin{table}[t]
\renewcommand{\arraystretch}{1.5}
\caption{Summary of Retention time for 8T SRAM augmented bit-cell}
\label{tab:tab1}
\begin{center}
 \begin{tabular}{|c|c|c|c|c|} 
 \hline
 Temperature & VWL1 & VWL2 & Retention time \\ [0.75ex] 
 \hline
85\degree C & -0.1V & 0V &~25 $\mu$s \\
\hline
25\degree C & 0V & 0V &~250 $\mu$s\\
 \hline
 25\degree C & -0.1V & 0V &  milli-sec\\
 \hline
\end{tabular}
\end{center}
\label{tab1}
\end{table}

\begin{table}[t]
\renewcommand{\arraystretch}{1.5}
\caption{Summary of Retention Time for 7T SRAM augmented bit-cell}
\label{tab:tab1}
\begin{center}
 \begin{tabular}{|c|c|c|c|} 
 \hline
 Temperature & VWL & Retention time \\ [0.75ex] 
 \hline
85\degree C & 0V & ~4 $\mu$s\\
\hline
25\degree C & 0V & \textgreater 50 $\mu$s\\
 \hline
\end{tabular}
\end{center}
\label{tab2}
\end{table}

\begin{table}[b]
\caption{Power/Energy consumption of 8T SRAM augmented cell}
\renewcommand{\arraystretch}{1.5}
\label{tab:tab1}
\begin{center}
 \begin{tabular}{|c|c|c|c|} 
 \hline
 Temperature & Region & 6T SRAM & 8T SRAM  \\ [0.75ex]
 \hline
%85\degree C& Hold & 44.2 nW & 1.9888uW & 1.4536uW \\ 
85\degree C& Hold & 0.448 uW & 0.603 uW  \\ 
 \hline
%85\degree C& Read & 3580 nW & 1.5693uW & 854.4679nW \\
 %\hline
 85\degree C& Read & 1.83 fJ & 3.37 fJ  \\
 \hline
 85\degree C&Write & 2.07 fJ & 8.32 fJ  \\
 \hline
 
 %25\degree C& Hold & 44.2 nW & 1.8069uW & 774.8993nW \\ 
 %\hline
%25\degree C& Read & 3580 nW & 1.5353uW & 619.3957nW \\
 %\hline
 %25\degree C&Write & 3462 nW & 13.4698nW & 870.3206nW \\
 %\hline
\end{tabular}
\end{center}
\label{tab3}
\end{table}

\begin{table}[b]
\caption{Power/Energy consumption of 7T SRAM augmented cell}
\renewcommand{\arraystretch}{1.5}

\label{tab:tab1}
\begin{center}
 \begin{tabular}{|c|c|c|c|c|} 
 \hline
 Temperature & Operation & 6T SRAM & 7T-Normal & 7T-AMC \\ [0.75ex] 
 \hline
 85\degree C & Hold & 0.448 uW & 0.430 uW & 0.59 uW \\ 
 \hline
 85\degree C & Read & 1.83 fJ & 3.53 fJ & 3.12 fJ \\
 \hline
 85\degree C & Write & 2.07 fJ & 2.02 fJ & 0.99 fJ \\
 \hline
\end{tabular}
\end{center}
\label{tab4}
\end{table}

Interestingly, augmented memory computing can be combined with in-memory computing techniques for additional benefits. Both analog and digital in-memory computing techniques have been presented in various previous works for static (SRAM) \cite{agrawal2017x} and dynamic (DRAM) bit-cells \cite{ali2019memory}. These in-memory techniques can be easily applied to the AMC bit-cells (specifically the 8T dual bit AMC cell) while operating in augmented computing mode. For example, the 8T dual bit-cell can be configured to store one SRAM-like and one DRAM-like data. During read operation multiple wordlines can be activated and digital or analog in-memory computing can be achieved while the 8T cell is operating in augmented memory mode. The FILO readout for 8T augmented bit-cell could still be enforced while performing in-memory computing operations to ensure the DRAM data is not inadvertently destroyed while accessing the SRAM data. Further, algorithm hardware co-design can be invoked to leverage trade-off between retention time, power consumption and end application accuracy \cite{liu2012raidr, giterman2019gc}. Augmented bit-cells thus provide multiple operational mode - 1) the Normal mode, 2) only Augmented computing mode, 3) only in-memory computing mode, and 4) Augmented + in-memory/near-memory computing mode. %Additionally, it is well-known that analog in-memory computing results in approximate computations. Thus, based on a specific application one could envision a memory system wherein as per the required end-application accuracy the memory array can  be operated in one among the available modes. 

\begin{table}[htbp]
\renewcommand{\arraystretch}{1.5}
\caption{Summary of Read and Write Delay for 8T SRAM augmented cell}
\label{tab:tab1}
\begin{center}
 \begin{tabular}{|c|c|} 
 \hline
  Write Delay & Read Delay \\ [0.75ex] 
 \hline
 ~1 ns &  ~15 ns\\
\hline
\end{tabular}
\end{center}
\label{tab5}
\end{table}

\begin{table}[htbp]
\renewcommand{\arraystretch}{1.5}
\caption{Summary of Read and Write Delay for 7T SRAM augmented cell}
\label{tab:tab1}
\begin{center}
 \begin{tabular}{|c|c|c|} 
 \hline
   & Write Delay & Read Delay \\ [0.75ex] 
 \hline
Data (0,0)  & ~0.4 ns &  ~0.4 ns\\
\hline
Data (0,1)/(1,0)  & ~0.5 ns & ~1.5 ns\\
\hline
\end{tabular}
\end{center}
\label{tab6}
\end{table}

%\begin{figure}[t]
%    \centering
%    \includegraphics[width=0.5\textwidth]{read_00.png}
%    \caption{The readout of data (0,0)}
%    \label{fig:figure6}
%\end{figure}

\section{Conclusion}

On-chip memory capacity is a key factor for many data intensive applications. In this paper, for the first time, we propose novel augmented memory bit-cells that can operate like conventional SRAM cells during normal mode of operation and can dynamically increase their storage capacity in the augmented mode of operations. We specifically present a 8 transistor SRAM cell that can store one SRAM-like and one DRAM-like bit, simultaneously, within the memory bit-cell. Similarly, our proposed 7 transistor SRAM bit-cell can store a ternary bit (three levels) in a dynamic fashion during the augmented mode of operation. Advantageously, the presented augmented bit-cells are amenable to in-memory compute paradigm that can provide added energy and throughput benefits. The functionality of the presented bit-cells has been confirmed by extensive simulations at Globalfoundries 22nm FD-SOI technology node. In summary, the concept of augmented memory bit-cells brings in a new dimension to accelerate data intensive application by dynamically augmenting the on-chip memory storage capacity.

\section*{Acknowledgement}
We thank Globalfoundries for providing the PDK and support for 22nm FDX technology.
 
\bibliographystyle{IEEEtran}
\bibliography{ref}

\end{document}